\documentclass[twocolumn,aps,prb,superscriptaddress,10pt]{revtex4}
\usepackage{amsmath}
\usepackage{amsfonts}
\usepackage{amssymb}
\usepackage{mathrsfs}
\usepackage{graphicx}
\usepackage{epstopdf}
\usepackage{subfigure}
\usepackage[colorlinks,
            linkcolor=blue,
            anchorcolor=blue,
            citecolor=blue,
            urlcolor=blue
            ]{hyperref}
\usepackage{txfonts}%

\providecommand{\U}[1]{\protect\rule{.1in}{.1in}}

\begin{document}
\title{Electronic structure and coexisting topological states in ferromagnetic and antiferromagnetic phases of $\mathbf{MnBi_{2}Te_{4}}$ quantum wires}
\author{Jian Li} 
\email{jianli@cqupt.edu.cn}
\affiliation{School of Science, Chongqing University of Posts and Telecommunications, Chongqing 400065, China}
\affiliation{Institute for Advanced Sciences, Chongqing University of Posts and Telecommunications, Chongqing 400065, China}
\affiliation{Southwest Center for Theoretical Physics, Chongqing University, Chongqing 401331, China}
\author{Zhu-Cai Yin} 
\affiliation{School of Science, Chongqing University of Posts and Telecommunications, Chongqing 400065, China}
\affiliation{Institute for Advanced Sciences, Chongqing University of Posts and Telecommunications, Chongqing 400065, China}
\author{Qing-Xu Li} 
\affiliation{School of Science, Chongqing University of Posts and Telecommunications, Chongqing 400065, China}
\affiliation{Institute for Advanced Sciences, Chongqing University of Posts and Telecommunications, Chongqing 400065, China}
\author{Jia-Ji Zhu } 
\email{zhujj@cqupt.edu.cn}
\affiliation{School of Science, Chongqing University of Posts and Telecommunications, Chongqing 400065, China}
\affiliation{Institute for Advanced Sciences, Chongqing University of Posts and Telecommunications, Chongqing 400065, China}
\affiliation{Southwest Center for Theoretical Physics, Chongqing University, Chongqing 401331, China}
\date{\today}

\begin{abstract}
We theoretically investigate the electronic structure of cylindrical magnetic topological insulator quantum wires in $\mathrm{MnBi_{2}Te_{4}}$. Our study reveals the emergence of topological surface states in the ferromagnetic phase, characterized by spin-polarized subbands resulting from intrinsic magnetization. In the antiferromagnetic phase, we identify the coexistence of three distinct types of topological states, encompassing both surface states and central states.
\end{abstract}
\maketitle



\section{\label{sec:level1}Introduction}
The interaction between nontrivial topological states and the magnetic order of magnetic topological insulators (MTI) gives rise to unique physical properties, such as the quantum anomalous Hall effect~\cite{chang2023colloquium, tokura202310} and axion insulators~\cite{li2024high,zhuo2023axion,lin2024spin,li2024dissipationless}. These properties offer diverse potential applications, ranging from low-power topological spintronic devices to topological quantum computing~\cite{zhang2023magnetic,wang2024topological,lian2018topological}. Since its discovery, the first intrinsic magnetic topological insulator, $\mathrm{MnBi_{2}Te_{4}}$~\cite{li2024progress}, has attracted considerable attention, sparking intense research interest~\cite{Dong2023,Wang2023}.

The relentless evolution of molecular beam epitaxial techniques and etching processes has paved the way for the fabrication of isolated semiconductor heterostructures~ \cite{zhai2023epitaxial,Paul2023,PhysRevLett.130.057001}. These heterostructures enable charge carrier confinement in both two-dimensional and three-dimensional spaces, leading to intriguing effects in condensed matter physics~\cite{ashoori1996electrons,ansaloni2020single,yazdani2020charge,Liu2021,Jin_2024}. The confined electrons and holes in momentum space manifest as zero-dimensional (0D) structures for quantum dots (QDs) and one-dimensional (1D) structures for quantum wires (QWRs) ~\cite{septianto2023enabling,heedt2021shadow,Science.1996.271.933}.
Various semiempirical approaches, such as the $\mathbf{k\cdot p}$ method~\cite{PhysRevResearch.1.013001,Zhao2023}, self-consistent calculations based on density-functional theory~\cite{PhysRevB.47.4413, PhysRevLett.111.156402}, and the tight-binding approach~\cite{PhysRevB.108.235410,PhysRevB.108.035147,PhysRevB.108.085419}, have been utilized to determine the electronic structure of conventional semiconductor QWRs. Electrons in QWRs can move freely along only in one direction, and their density of states is concentrated in a few peaks. This unique property makes QWRs appropriate for various applications, including complementary-metal-oxide-semiconductor field-effect transistors~\cite{Nanotechnology.2020.31.445204}, nanowires-based field-effect transistors~\cite{ahmad2018recent}, and single-electron transistors~\cite{AppliedPhysicsLetters.2003.83.2052}. Additionally, the light-manipulating characteristics of QWRs could be exploited for creating highly effective solar energy devices~\cite{zhang2022large}.

Recently, numerous varieties of QWRs have been proposed, including superconducting QWRs~\cite{Science.2014.344.623}, metal QWRs ~\cite{tsai2022antisite}, insulator QWRs ~\cite{PhysRevB.81.113302,vaitiekenas2021zero}, and topological insulator (TI) QWRs~\cite{legg2022metallization,JournalofAppliedPhysics.2011.110.093714}. Among these, MTI QWRs are particularly noteworthy. In the quantum anomalous Hall phase, MTI/superconductor QWRs demonstrate significant stability ranges for both topological and non-topological states, enabling the creation and control of Majorana particles on the surface of the topological insulator. The construction of quasi-1D quantum anomalous Hall systems could enable a scalable approach to quantum computing~\cite{Phys.Rev.B.2018.97.104504}, potentially leading to the experimental observation of helical states in magnetically doped QWRs~\cite{Phys.Rev.B.2018.97.081102}. However, initial MTI QWRs depend on randomly distributed magnetic dopants, which results in highly inhomogeneous magnetic and electronic properties of these materials, limiting their applications in spintronics. The introduction of the layered van der Waals material $\mathrm{MnBi_{2}Te_{4}}$, a magnetic topological insulator, addresses these issues. $\mathrm{MnBi_{2}Te_{4}}$ exhibits either ferromagnetic or antiferromagnetic characteristics depending on whether the sample consists of an odd (even) number of single layers, respectively, giving rise to zero-field quantum anomalous Hall  (for odd layers) or intrinsic axion insulator (for even layers) states~\cite{Nature.2019.576.416, 2DMaterials.2017.4.025082, Phys.Rev.Lett.2019.122.107202,ScienceAdvances.2019.5.eaaw5685,Science.2020.367.895,Ji2021}. The $\mathrm{MnBi_{2}Te_{4}}$ QWRs could effectively resolve the issues found in the initial MTI QWRs, facilitating the pursuit of spintronics-related technologies.

In this work, we focus on the distinctive features of the single electron energy spectrum in the cylindrical ferromagnetic (FM) and antiferromagnetic (AFM) phases of $\mathrm{MnBi_{2}Te_{4}}$ QWRs individually, which is crucial for comprehending the few-electron characteristics and charge transport properties within these systems. For the FM phase, we predict a surface state where the spin degeneracy is broken by the intrinsic magnetic properties present within the bulk band gap, in contrast with TI QWRs. For the AFM phase, our numerical results indicate that three subbands are present within the bulk band gap of the $\mathrm{MnBi_{2}Te_{4}}$ QWRs, characterized by the coexistence of surface states near the boundary and central states near the center of the QWRs.

\section{\label{sec:level2} Model and method}
\begin{figure}[ptbh]
\includegraphics[width=0.46\textwidth]{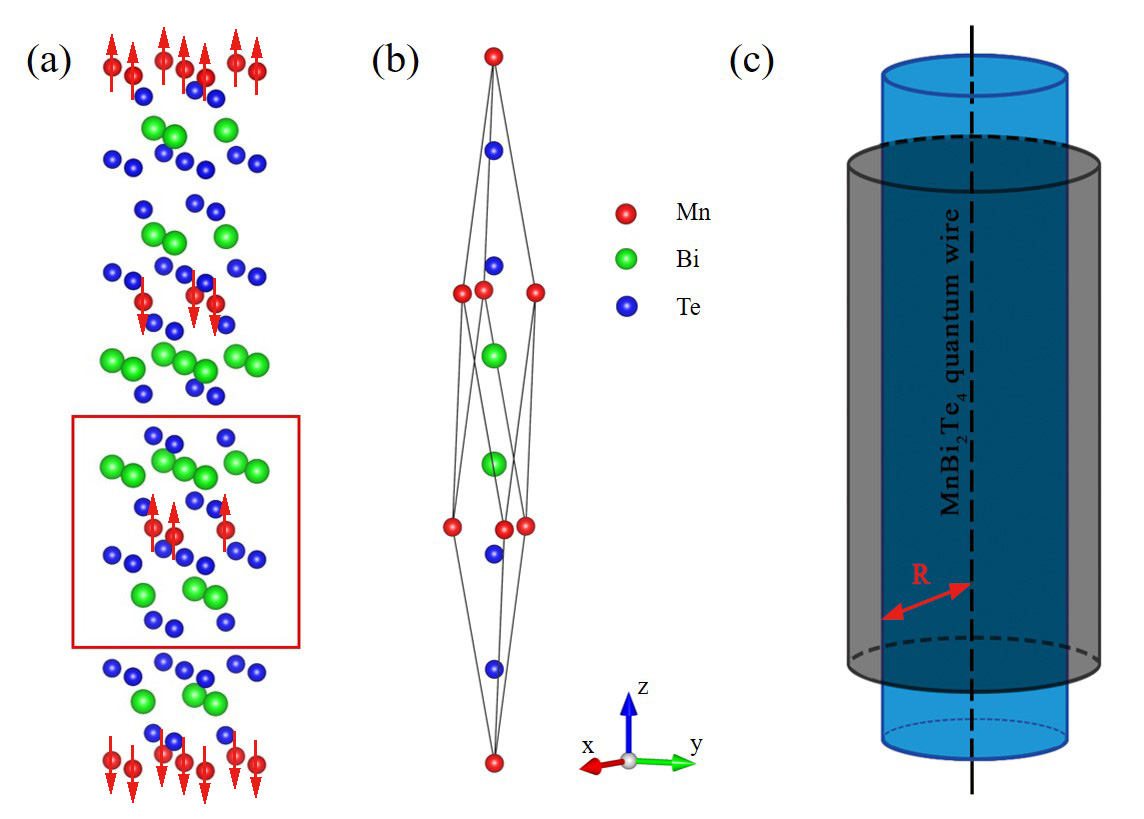}
\caption{(Color online) (a) The crystal structure of $\mathrm{MnBi_{2}Te_{4}}$, showing the seven-layer stacked structure of Ti-Bi-Te-Mn-Te-Bi-Te highlighted in the red box, where the red arrows depict the spin moment of the Mn atom. (b) A schematic diagram of the paramagnetic rhombohedral unit cell of the bulk $\mathrm{MnBi_{2}Te_{4}}$ . (c) A schematic diagram of $\mathrm{MnBi_{2}Te_{4}}$ QWR oriented along the $z$-axis, where $R$ is the radius of the QWR.}
\label{fig:1}%
\end{figure}

We consider a cylindrical $\mathrm{MnBi_{2}Te_{4}}$ QWRs as schematically illustrated in FIG.~\ref{fig:1}(c). The low-energy properties of $\mathrm{MnBi_{2}Te_{4}}$ QWRs near the $\Gamma$ point can be effectively described by a four-band effective Hamiltonian~\cite{Phys.Rev.Lett.2019.122.206401} with lateral confinement in the basis states $\left \vert P1_{z}^{+},\uparrow \right \rangle $, $\left \vert P2_{z}^{-},\uparrow \right \rangle $, $\left \vert P1_{z}^{+},\downarrow \right \rangle $ and $\left \vert P2_{z}^{-},\downarrow \right \rangle $. Here, the superscripts $``+"$ and $``-"$ denote the parity of the corresponding states, $P$ represents the $p$-like orbital states of the atoms, where 1(2) corresponds to Bi (Te) atoms, and $\uparrow(\downarrow)$ signifying the spin-up (spin-down) state, respectively.

It has been proposed that the magnetic ground state of $\mathrm{MnBi_{2}Te_{4}}$ is an AFM topological insulator state. Additionally, its magnetic states can be tuned from AFM to FM under the influence of external magnetic fields.  Consequently,  we explore the physical properties of $\mathrm{MnBi_{2}Te_{4}}$ QWRs in both the FM and AFM phases. The etching process generates a sharp confinement potential $V\left(\rho \right)$ along the radial direction of the QWRs. For simplicity, we take $V\left(\rho \right)$ as a cylindrical hard wall potential, \textit{i.e.}, $V\left(\rho \right)=0$ for $\rho<R$ and $V\left(  \rho \right)  =\infty$ for $\rho>R$.

The low-energy effective Hamiltonian of FM $\mathrm{MnBi_{2}Te_{4}}$ QWRs is
\begin{equation}
 H_{FM}^{QWRs}\left(k\right)=H_{N}\left(k\right)+\delta H_{FM}\left(
k\right)+V\left(\rho \right)
\label{FM},
\end{equation}
where $H_{N}\left(k\right)$ is the nonmagnetic state Hamiltonian:
\begin{equation}
H_{N}(k)=\varepsilon_{0}(k)+\left(
\begin{array}
[c]{cccc}%
M_{\gamma}(k) & A_{1}k_{z} & 0 & A_{2}k_{-}\\
A_{1}k_{z} & -M_{\gamma}(k) & A_{2}k_{-} & 0\\
0 & A_{2}k_{+} & M_{\gamma}(k) & -A_{1}k_{z}\\
A_{2}k_{+} & 0 & -A_{1}k_{z} & -M_{\gamma}(k)
\end{array}
\right)
\label{eq:one},
\end{equation}
where, $k_{\pm}=k_{x}\pm ik_{y}$, $\varepsilon_{0}(k)=C+D_{1}k_{z}^{2}+D_{2}\left(
k_{x}^{2}+k_{y}^{2}\right)$, $M_{\gamma}(k)=M_{0}^{\gamma}+B_{1}^{\gamma}k_{z}^{2}+B_{2}^{\gamma}(k_{x}%
^{2}+k_{y}^{2})$, the band parameters are taken from Ref. [\onlinecite{Phys.Rev.Lett.2019.122.206401}]: $C=0.0539~\mathrm{eV}$, $D_{1}=-4.0339~\mathrm{eV\mathring{A}^{2}}$, $D_{2}=4.4250~\mathrm{eV\mathring{A}%
^{2}}$, $A_{1}=1.3658~\mathrm{eV\mathring{A}}$, $A_{2}=1.9985~\mathrm{eV\mathring{A}}$, $M_{0}^{\gamma
}=0.0307~\mathrm{eV}$, $B_{1}^{\gamma}=4.2857~\mathrm{eV\mathring{A}^{2}}$, $B_{2}^{\gamma}=8.3750~\mathrm{eV\mathring
{A}^{2}}$. And $\delta H_{FM}\left(k\right)$ is the FM perturbative term given by :
\begin{equation}
\delta H_{FM}(k)=\left(
\begin{array}
[c]{cccc}%
M_{1}(k) & A_{3}k_{z} & 0 & A_{4}k_{-}\\
A_{3}k_{z} & M_{2}(k) & -A_{4}k_{-} & 0\\
0 & -A_{4}k_{+} & -M_{1}(k) & A_{3}k_{z}\\
A_{4}k_{+} & 0 & A_{3}k_{z} & -M_{2}(k)
\end{array}
\right)
\label{eq:two},
\end{equation}
where $M_{1,2}(k)=M_{\alpha}(k)\pm M_{\beta}(k)$ characterize the Zeeman coupling with the magnetized Mn orbitals, and $M_{j}(k)=M_{0}^{j}+B_{1}^{j}k_{z}^{2}+B_{2}^{j}(k_{x}^{2}+k_{y}^{2})$, with  $j=\alpha,\beta$. The band parameters are also taken from  Ref. [\onlinecite{Phys.Rev.Lett.2019.122.206401}]: $A_{3}=-0.5450~\mathrm{eV\mathring{A}}$, $A_{4}=1.1384~\mathrm{eV\mathring{A}}$, $M_{0}^{\alpha}=-0.1078~\mathrm{eV}$, $B_{1}^{\alpha}=0.6232~\mathrm{eV\mathring{A}^{2}}$, $B_{2}^{\alpha}=0.6964~\mathrm{eV\mathring{A}^{2}}$, $M_{0}^{\beta}=-0.0880~\mathrm{eV}$, $B_{1}^{\beta}=2.7678~\mathrm{eV\mathring{A}^{2}}$, $B_{2}^{\beta}=1.9650~\mathrm{eV\mathring{A}^{2}}$.

For the AFM phase $\mathrm{MnBi_{2}Te_{4}}$ QWRs, the effective Hamiltonian is
\begin{equation}
H_{AFM}^{QWRs}\left(  k\right)  = H_{AFM}\left(
k\right)  +V\left(  \rho \right)
\end{equation}
where
\begin{equation}
H_{AFM}(k)=\varepsilon_{0}^{\prime}(k)+\left(
\begin{array}
[c]{cccc}%
M^{\prime}(k) & A_{1}^{\prime}k_{z} & 0 & A_{2}^{\prime}k_{-}\\
A_{1}^{\prime}k_{z} & -M_{\gamma}^{\prime}(k) & A_{2}^{\prime}k_{-} & 0\\
0 & A_{2}^{\prime}k_{+} & M_{\gamma}^{\prime}(k) & -A_{1}^{\prime}k_{z}\\
A_{2}^{\prime}k_{+} & 0 & -A_{1}^{\prime}k_{z} & -M_{\gamma}^{\prime}(k)
\end{array}
\right)
\label{eq:four},
\end{equation}
the Hamiltonian of $H_{AFM}(k)$ is identical to that of the nonmagnetic form  $H_{N}\left(  k\right)$, i.e. $\varepsilon_{0}^{\prime}(k)=C^{\prime}+D_{1}^{\prime}k_{z}^{2}+D_{2}^{\prime}\left(  k_{x}^{2}+k_{y}^{2}\right)$, $M^{\prime}(k)=M_{0}^{\prime}+B_{1}^{\prime}k_{z}^{2}+B_{2}^{\prime}(k_{x}%
^{2}+k_{y}^{2})$, but with different band parameters~\cite{Phys.Rev.Lett.2019.122.206401}, $C^{\prime}=-0.0048~\mathrm{eV}$, $D_{1}^{\prime}=2.7232~\mathrm{eV\mathring{A}^{2}}$, $D_{2}^{\prime}=17.0000~\mathrm{eV\mathring{A}^{2}}$, $A_{1}^{\prime}=2.7023~\mathrm{eV\mathring{A}}$, $A_{2}^{\prime}=3.1964~\mathrm{eV\mathring{A}}$, $M_{0}^{\prime}=-0.1165~\mathrm{eV}$, $B_{1}^{\prime}=11.9048~\mathrm{eV\mathring{A}^{2}}$, $B_{2}^{\prime}=9.4048~\mathrm{eV\mathring{A}^{2}}$.

The single electron states and the corresponding eigenstates of the QWRs are derived from the Schr\"{o}dinger equation:
\begin{equation}
H_{FM/AFM}^{QWRs}\left( k\right) \left\vert \Psi \left( k\right) \right\rangle
=E_{FM/AFM}^{k}\left\vert \Psi \left( k\right) \right\rangle
\label{schrodinger},
\end{equation}
which can be determined numerically by expanding the wave function in terms of the cylindrical Bessel functions of the first kind~\cite{Phys.Rev.B.2014.90.115303}:
\begin{equation}
\Psi =\sum\limits_{n,m}\left(
\begin{array}{c}
C_{n,m}^{(1)}\varphi _{n,m} \varphi _{z}\\
C_{n,m-1}^{(2)}\varphi _{n,m-1} \varphi _{z}\\
C_{n,m}^{(3)}\varphi _{n,m} \varphi _{z}\\
C_{n,m+1}^{(4)}\varphi _{n,m+1}\varphi _{z}%
\end{array}%
\right) ,  \label{eq:three}
\end{equation}
where $\varphi _{n,m}=\frac{1}{\sqrt{\pi }\times R\times J_{m+1}\left( k_{n}^{m}\right) }J_{m}\left( k_{n}^{m}\rho /R\right) e^{im \theta }$, $\varphi _{z}=\frac{1}{\sqrt{2\pi }}e^{ik_{z}z}$,  $C_{n,m}^{(i)}$ are the expanding coefficient. The number of eigenstates used in the expansion is selected to guarantee the convergence of the calculated energies of the QWRs within and nearby the bulk gap. $\hat{L}_{z}=-i\partial/\partial \theta$ is the angular momentum of the envelope function, therefore,  $m= \left\langle \hat{L}_{z}\right\rangle =0,\pm1,\pm2\cdots$ is a good quantum number due to therotational symmetry of $H_{FM}^{QWRs}\left(  k\right)  $  ( or $H_{AFM}^{QWRs}\left(  k\right)  $) around the $z$ axis,  $k_{n}^{m}$ is the $n$th zero point of the first class of $m$th order Bessel function $J_{m}(x)$, $k_{z}$ is the wavevector along the QWRs direction, i.e., the $z$ axis, and $R$ is the radius of the cylindrical QWRs.

To obtain the single-electron states, we left-multiply the secular equation (\ref{schrodinger}) by $\left( \varphi _{n^{\prime },m^{\prime }}^{\ast }\varphi _{z}^{\ast },\varphi _{n^{\prime },m^{\prime }-1}^{\ast }\varphi _{z}^{\ast },\varphi _{n^{\prime },m^{\prime }}^{\ast}\varphi _{z}^{\ast },\varphi {n^{\prime },m^{\prime }+1}^{\ast }\varphi {z}^{\ast }\right) $ to obtain the eigenvalues and expansion coefficients $C{n,m}^{(i)}$ by solving the eigenvalue problem. The basis set $\varphi_{n,m}(\rho ,  \theta)$ and $\varphi _{z} $  obeys the orthonormalized condition $\int_{0}^{2\pi }\int_{0}^{R}\varphi _{n,m}(\rho, \theta )\varphi _{n^{\prime },m}(\rho, \theta )\rho d\rho d\theta =\delta _{n,n^{\prime }}\delta _{m,m^{\prime }}$ and  $\int_{-\infty }^{\infty }\varphi _{z}\varphi _{z}^{\ast }dz=\int_{-\infty }^{\infty }\frac{1}{2\pi }e^{i\left( k_{z}-k_{z}^{^{\prime }}\right) z}dz=\delta \left( k_{z}-k_{z}^{^{\prime }}\right) $.

\section{Results and discussions}
In the first subsection, we showcase the electronic structure and density of states (DOS) of cylindrical $\mathrm{MnBi_{2}Te_{4}}$ QWRs in its FM phase. In this phase, surface states are found to exist within the bulk band gap, and the spectra of the surface states exhibit non-degenerate and spin polarization due to the broken $ \mathcal{T} \mathcal{I}$ symmetry, where  $\mathcal{T} $ is time reversal symmetry, $\mathcal{I}$ is inversion symmetry.

In the second subsection, we present the electronic structure and DOS of the cylindrical $\mathrm{MnBi_{2}Te_{4}}$ QWRs in their AFM phase. This phase exhibits coexisting states within the bulk band gap. The energy spectrum shows two-fold degeneracy, which arises from the conservation of $\mathcal{T}  \mathcal{I}$ symmetry.

\subsection{\label{sec:level3} Ferromagnetic phase}
\begin{figure}[ptbh]
\includegraphics[width=0.46\textwidth]{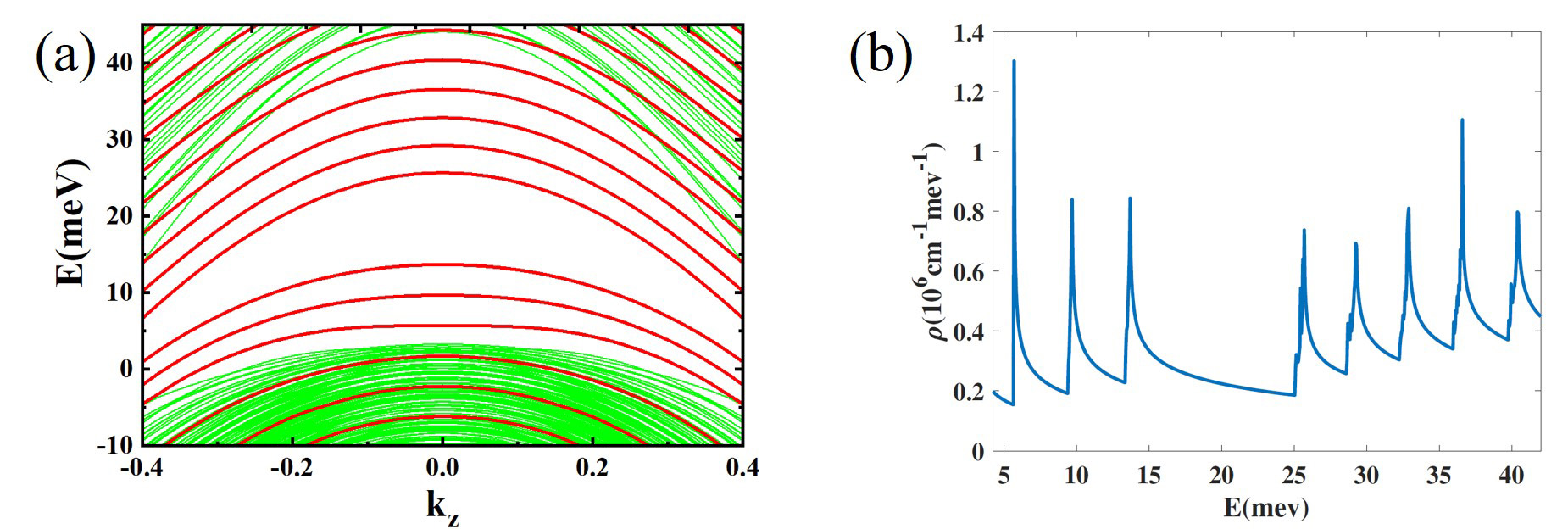}
\caption{(Color online) (a) Energy spectrum of the FM $\mathrm{MnBi_{2}Te_{4}}$ QWRs with radius $R = 60 ~\mathrm{nm}$. The red solid lines represent the surface states, while the green solid lines represent the bulk states. (b) DOS of the surface states of the QWRs within the bulk band gap.}%
\label{fig:2}%
\end{figure}

To begin, we present the energy spectrum of the FM phase of $\mathrm{MnBi_{2}Te_{4}}$ QWRs in Fig.~\ref{fig:2}(a), calculated numerically using the four-band $\mathbf{k \cdot p}$ model. The surface states possess a mini-gap (approximately $8\ \mathrm{meV}$) between the up- and down-branches of the surface state subbands (depicted by the red solid line in Fig.~\ref{fig:2}(a)), arising from the quantization of orbital angular momentum. Additionally, the energy spectrum of the surface states is fully spin-polarized.

From the subband dispersions of the surface states of the QWRs shown in Fig.~\ref{fig:2}(a), one can calculate the DOS per unit length by utilizing the following equation:
\begin{equation}
\rho^{1D}(E)=\sum \limits_{i=1}^{n}\left(  \frac{2m^{\ast}}{\hbar^{2}}\right)
^{\frac{1}{2}}\frac{1}{\pi \left(  E-E_{i}\right)  ^{\frac{1}{2}}}\Theta \left(
E-E_{i}\right)
\label{eq:five},
\end{equation}
where $\Theta$ represents the unit step function, and the summation over $i$ includes all subbands. Fig.~\ref{fig:2}(b) illustrates the DOS of the surface states of the QWRs. The peaks in the DOS, which occur near the subband maxima, stem from the 1D nature of the surface states subbands. Consequently, charge accumulation is expected to occur around these points.

\begin{figure}[ptbh]
\includegraphics[width=0.46\textwidth]{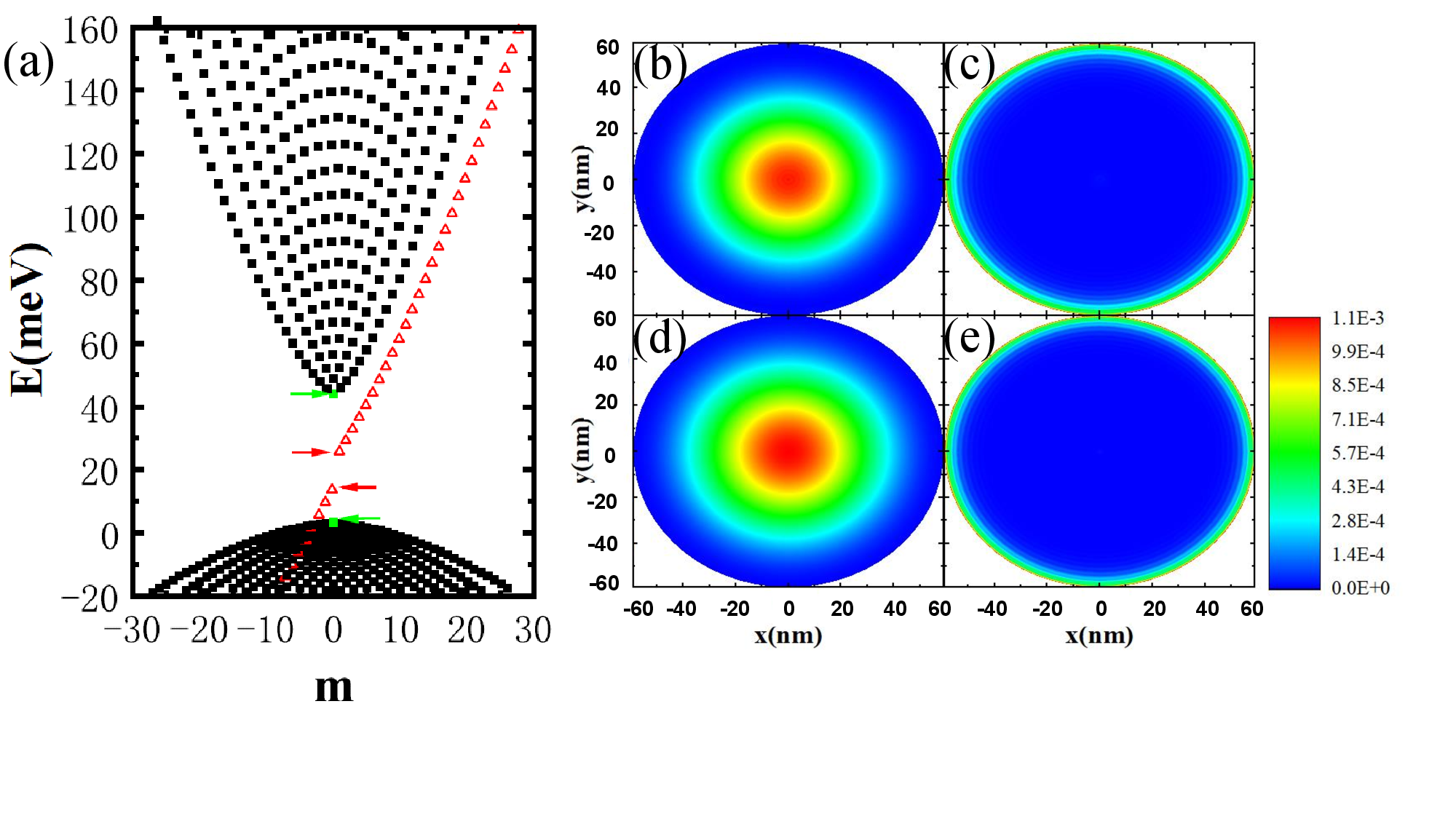}
\caption{(Color online)
(a) Energy spectrum of the FM phase of $MnBi_{2}Te_{4}$ QWRs is presented as a function of the angular momentum quantum number $m$ with a radius $R = 60 ~\mathrm{nm}$ and $k_{z}=0$. In (b) and (d), the density distributions of the lowest conduction band and highest valence band states are shown, marked by green arrows in (a). Panels (c) and (e) display the density distributions of a specific state among the surface states, indicated by red arrows in (a).}
\label{fig:3}%
\end{figure}

In Fig.~\ref{fig:3}, we plot the energy spectra of the surface and bulk states of the FM phase of $\mathrm{MnBi_{2}Te_{4}}$ QWRs, with respect to the angular momentum quantum number $m$ at the Brillouin zone's (BZ) center where $k_{z}=0$. The spectra of the surface states exhibit a fully spin-polarized characteristic and approximately adhere to a linear dispersion in relation to $m$ in Fig.~\ref{fig:3}(a). Notably, in contrast to the TI cylindrical QWRs reported in Ref.~\onlinecite{JournalofAppliedPhysics.2011.110.093714}, only one spin branch is present here due to the broken  $\mathcal{T}  \mathcal{I}$  symmetry. The spins of electrons conducting electrical current are predominantly aligned along the magnetization direction, which holds significant importance for spintronics applications.

The density distributions of the bulk states for the conduction band and valence band are concentrated at the center of the QWRs, as illustrated in Fig.~\ref{fig:3}(b) and  \ref{fig:3}(d) respectively. In contrast to the bulk states, the surface states exhibit an interesting ring-like density distribution in the vicinity of the outer surface of the QWRs, as demonstrated in Fig.~\ref{fig:3}(c) and  \ref{fig:3}(e) respectively. When this massless Dirac fermion is restricted within the FM phase of $\mathrm{MnBi_{2}Te_{4}}$ QWRs, its lowest energy modes ought to be the \textquotedblleft whispering gallery \textquotedblright mode, analogous to an electron (hole) confined in TI cylindrical QWRs~\cite{JournalofAppliedPhysics.2011.110.093714} or QDs~\cite{Phys.Rev.Lett.2011.106.206802}.

\subsection{\label{sec:level4} Antiferromagnetic phase}

\begin{figure}[ptbh]
\includegraphics[width=0.46\textwidth]{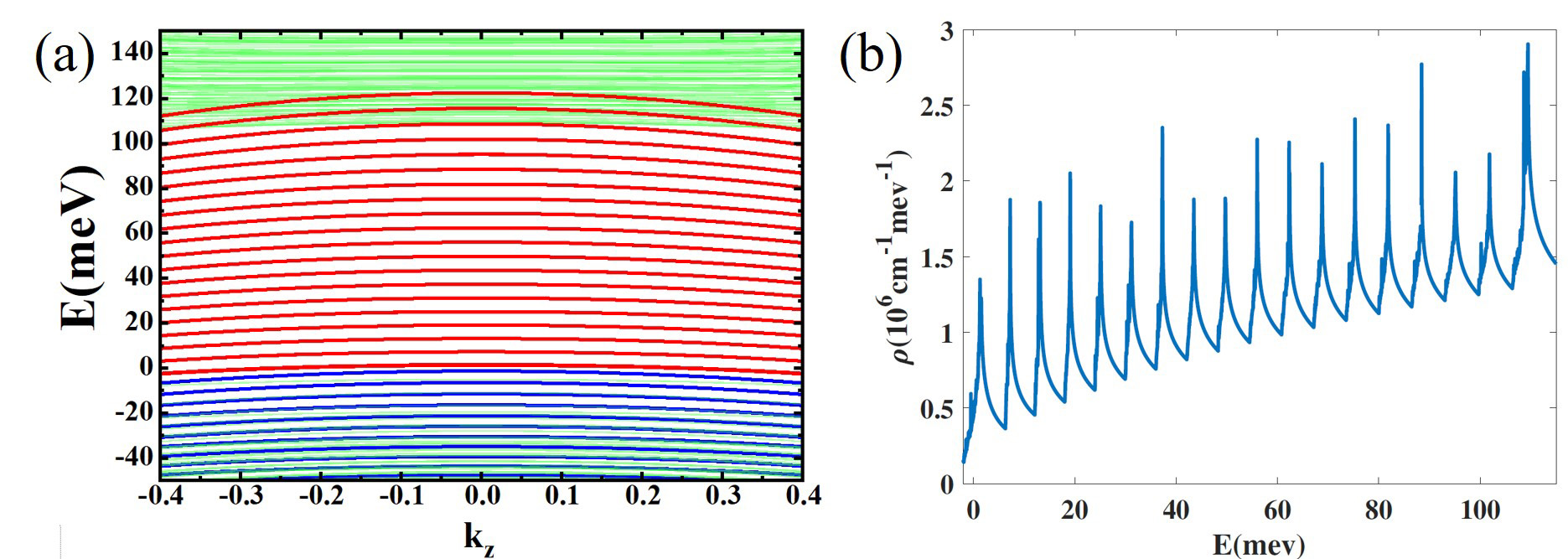}
\caption{(Color online) (a) Energy spectrum of the  AFM $\mathrm{MnBi_{2}Te_{4}}$ QWRs with a radius of $R = 60 ~\mathrm{nm}$. The red and blue solid lines represent the coexistent states, while the green solid lines signify the bulk states. (b) DOS of the coexistent states of the QWRs within the bulk band gap.}
\label{fig:5}%
\end{figure}

In the following sections, we delve into the electronic structure of the AFM phase of $\mathrm{MnBi_{2}Te_{4}}$ QWRs. The energy spectra of the surface and bulk states of the AFM phase are plotted as a function of the wavevector $k_{z}$ in Fig.\ref{fig:5}(a). In contrast to the FM phase of $\mathrm{MnBi_{2}Te_{4}}$ QWRs, the subbands (red and blue solid line in Fig.~\ref{fig:5}(a)) within the bulk band gap are not solely surface states but are rather coexisting states. Similar to the FM form, it is also possible to obtain the  DOS  per unit length of AFM phase  $\mathrm{MnBi_{2}Te_{4}}$ QWRs (see Fig.~\ref{fig:5}(b)). The peaks in the DOS around the subbands' maxima result from the one-dimensional characteristic of the coexistent states subbands. A larger DOS is observed due to spin degeneracy in comparison to the FM phase of $\mathrm{MnBi_{2}Te_{4}}$ QWRs. The prominent peak near the maxima in the DOS is ascribed to the flatness of the subbands.

\begin{figure}[ptbh]
\includegraphics[width=0.46\textwidth]{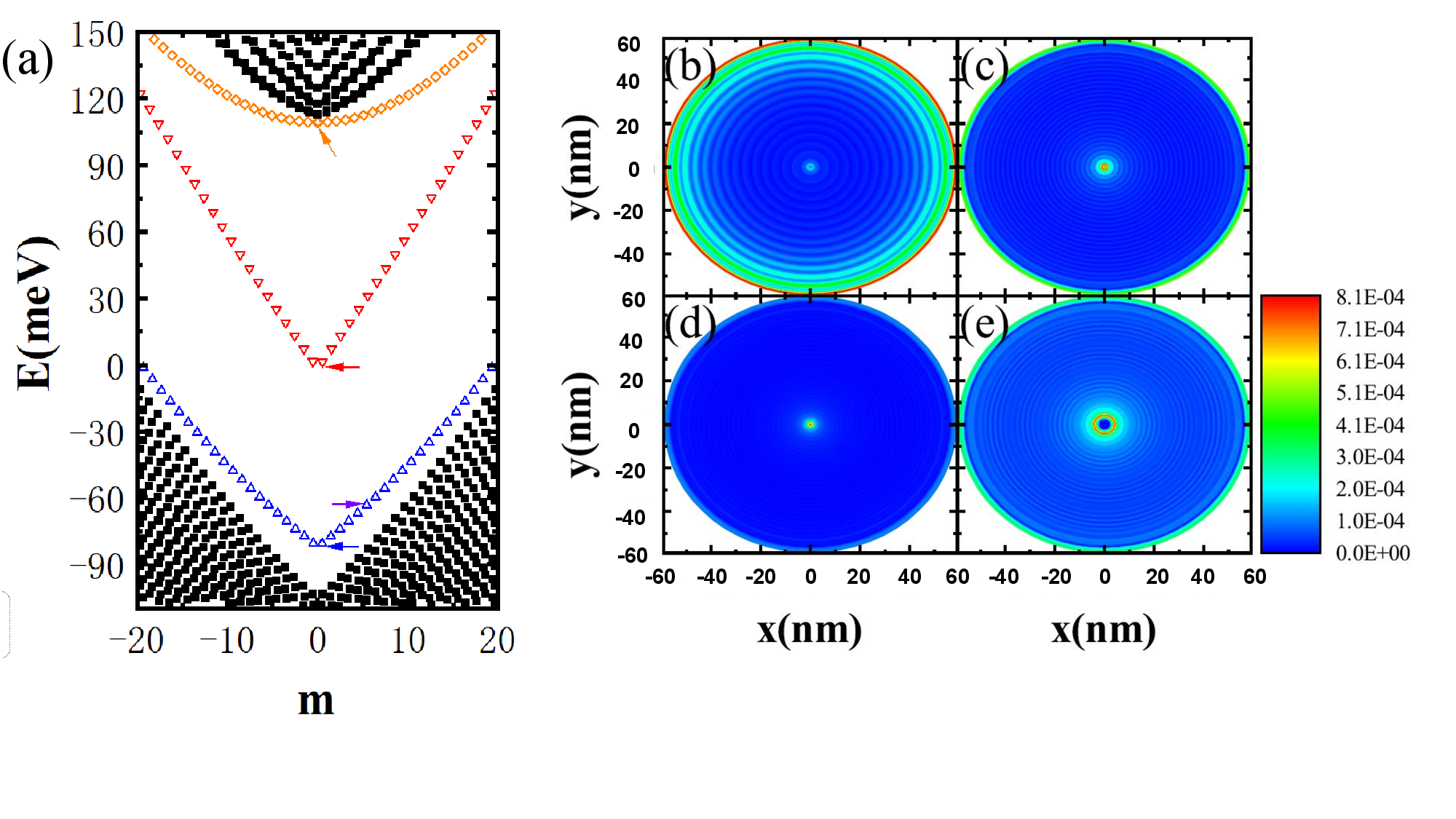}
\caption{(Color online)
(a) Energy spectrum of the AFM phase of $MnBi_{2}Te_{4}$ QWRs as a function of the angular momentum quantum number $m$ with a radius $R = 60 ~\mathrm{nm}$ and $k_{z}=0$.  Panels (b), (c), and (d) show the density distributions of the three coexisting states in the band gap (marked by yellow, red and blue arrows in (a) for $m=0$, separately). (e) The density distributions of the lowest coexistent states within the band gap (marked by green arrows in (a) for $m=5$).}
\label{fig:6}%
\end{figure}

To further clarify the electronic states of the AFM phase of $\mathrm{MnBi_{2}Te_{4}}$ QWRs, we present their energy spectra of the coexistent states and bulk states as a function of the angular momentum quantum number $m$ at the center of the BZ, $k_{z}=0$, in Fig.~\ref{fig:6}(a). It can be observed that three distinct subbands exist between the energy gaps, and the middle subband exhibits a linear dispersion with respect to $m$. Additionally, we have provided the density distribution of the three coexistent states between the energy gaps. The density distribution of these three distinct subband states reveal the coexistence of surface states near the boundary and centre states near the center of the QWRs, as shown in Fig.~\ref{fig:6}(b), (c), (d) and (e). It should be noted that although the surface state component of the lowest branch (marked by blue arrows in Fig.~\ref{fig:6}(a)) at angular momentum $m=0$ (see Fig.~\ref{fig:6}(d)) is not as obvious as at larger $m$ (see Fig.~\ref{fig:6}(e) for $m=5$ as an example), we still consider it to be part of the coexisting states overall. The novel coexistent states are significantly different from the surface states of $\mathrm{Bi_{2}Se_{3}}$ QWRs and those of previous FM QWRs. They can be manipulated by tuning the band parameter $D_{2}^{\prime}$ in $\varepsilon_{0}^{\prime}(k)$, which represents the kinetic energy of the $\mathbf{k \cdot p}$ Hamiltonian (\ref{eq:four}) in our calculation. Consequently, these states can be controlled via an external field.

\section{\label{sec:level5}Conclusions}
In summary, we investigated the electronic structure of $\mathrm{MnBi_{2}Te_{4}}$ QWRs in both the FM and AFM phases. For the FM phase QWRs, the numerical results reveal the presence of surface states within the bulk band gap. These surface states display spectra that are completely spin-polarized, meaning only one spin branch is present. For AFM phase QWRs, we identified the presence of three coexistent states within the energy gaps. Importantly, the behavior of the coexistent states can be controlled by an external field. This exploration of the electronic properties of $\mathrm{MnBi_{2}Te_{4}}$ QWRs in both FM and AFM phases provides valuable insights for potential applications in nanoelectronic devices.

\begin{acknowledgments}
This work has been sponsored by Natural Science Foundation of Chongqing, China (Grant No. CSTB2024NSCQ-MSX0736), Special Project of Chongqing Technology Innovation and Application Development (Major Project)(Grant No. CSTB2024TIAD-STX0035), and the research foundation of Institute for Advanced Sciences of CQUPT (Grant No. E011A2022328).
\end{acknowledgments}

\bibliographystyle{iopart-num}
\bibliography{MBTQWRS}
\end{document}